\documentclass[twocolumn,floats,showpacs,prb]{revtex4}
\usepackage{amsmath}
\usepackage{graphicx}

\begin{document}

\title{A Real Space Glue for Cuprate Superconductors}
\author{Xiuqing Huang$^{1,2}$}
\email{xqhuang@nju.edu.cn}
\affiliation{$^1$Department of Physics and National Laboratory of Solid State
Microstructure, Nanjing University, Nanjing 210093, China \\
$^{2}$ Department of Telecommunications Engineering ICE, PLAUST, Nanjing
210016, China}
\date{\today}

\begin{abstract}
In a recent article [Science \textbf{317}, 1705 (2007)], Anderson pointed
out that many theories about electron pairing in cuprate superconductors may
be on the wrong track and there is no reason to believe that the dynamic
screening (\textbf{k}-space) can provide a valid \textquotedblleft
glue\textquotedblright\ to hold the electron pairs together. On the other
hand, the most recent experimental observations imply the possible generic
existence of the real space localized Cooper pairs in amorphous insulating
and other nonsuperconducting systems [M. D. Stewart Jr. \textit{et al.},
Science \textbf{318}, 1273 (2007)]. It is therefore clear that the
\textquotedblleft glue\textquotedblright\ for high-$T_{c}$ superconductors
is relevant to the real space correlations. In this paper, we argue that
real space electron-electron interactions can play the role of
\textquotedblleft glue\textquotedblright\ in high-$T_{c}$ superconductors.
It is found that two localized electrons, due to a real space Coulomb
confinement effect, can be in pairing inside a single plaquette of the CuO
plane. The scenario suggests appearance of a dominating $d$-wave-like
pairing symmetry in the hole-doped cuprates, while a more complex mixed $%
(s+d)$ symmetry in the electron-doped cuprates. Based on the mechanism, the
relationships between the superconductivity and the charge-stripe order are
discussed. In the La$_{2-x}$Sr$_{x}$CuO$_{4}$ (LSCO) with $1/18\leq x\leq
1/4,$ we show that the paired electrons can self-organize into the dimerized
Wigner crystal. Two kinds of quasi-one-dimensional metallic magnetic-charge
stripes, where the superconductivity and dynamical spin density wave (SDW)
coexist, are analytically determined. Furthermore, the physical original of
the magic doping fractions (at $x=1/4$, 1/8, 1/9, 1/16 and 1/18) of
superconductivity and an analytical phase diagram for LSCO are given.
\end{abstract}

\maketitle

\section{Introduction}

It has been over two decades since the first discovery of high-$T_{c}$
superconductivity (HTSC) in copper oxide materials \cite{bednorz}.
Theoretically, soon after the discovery of the high-$T_{c}$ cuprates, the
resonating valence bond (RVB) theory, a strong coupling version of the spin
fluctuation approach, was introduced by Anderson \cite{anderson}. After
Anderson's original conjecture, Zhang and Rice derived a single-band
effective Hamiltonian for HTSC cuprates \cite{zhangfc}. Wen and Lee
developed the SU(2)-gauge theory of the strong coupling nature for the doped
$t$-$J$ model \cite{wen}. Varma and co-workers established the marginal
Fermi liquid theory to explain many of the anomalous behaviors in cuprates
\cite{varma}. Zhang proposed a very interesting and unified theory based on
the SO(5) symmetry of superconducting (SC) and antiferromagnetic (AF) phases
\cite{zhang}. Moreover, Dagotto \cite{dagotto} and Lee \textit{et al}. \cite%
{lee} gave comprehensive reviews of the mechanisms of the HTSC. In spite of
extensive theoretical efforts, the mechanism of the HTSC in the cuprates
remains a profound unsolved mystery of modern physics. Now, the majority
view on the mechanism is that the weak-coupling mean-field BCS theory \cite%
{bcs}, which is based on electron-phonon interaction, may not be suitable
for the high superconducting transition caused by strong correlations.
Experimentally, accumulating evidences show that the pairing mechanism of
HTSC is relevant to the real-space (a localized picture) electron-electron
correlations. One of the most striking experimental fact is the observation
of a periodic pattern in the local electron density of states (LDOS) by high
resolution scanning tunneling microscopy (STM) \cite{hanaguri}. Furthermore,
one of the most exciting experimental evidences in favor of the localized
Cooper pairs have just been reported \cite{stewart}, the excellent
experiment implies the possible generic existence of the real space
superconducting pair correlations in amorphous insulating and other
nonsuperconducting systems such as underdoped high-$T_{c}$ superconductors
above $T_{c}$.

Very recently, Anderson pointed out that many theories about electron
pairing in cuprate superconductors may be on the wrong track \cite{anderson0}%
. In the paper, it is clearly shown that the framework of dynamic screening
(quasiparticle) is impossible to provide the \textquotedblleft
glue\textquotedblright\ for the high-$T_{c}$ copper oxide superconductors,
as the net interaction between electrons is always repulsive. If this is the
correct viewpoint, but what will be the \textquotedblleft
glue\textquotedblright\ binding the electrons in pairs?

Compared to conventional superconductors, besides their exceptionally high
superconducting transition temperatures, there are many anomalous physical
properties in high-$T_{c}$ cuprates. It appears to be generally accepted
that conventional BCS superconductors are characterized by a standard $s$%
-wave, while the symmetry of the high-$T_{c}$ superconducting order
parameter is believed to be $d$-wave for the hole-doped ($p$-type) cuprates
\cite{shen,tsuei,wollman}. However, for the electron-doped ($n$-type)
cuprates there is much controversy about whether the order parameter has a $%
s $-wave symmetry \cite{wu,alff}, or $d$-wave \cite%
{armitage,tsuei1,snezhko,ariando}, or a nonmonotonic $d$-wave \cite{matsui},
or a crossover from a $d$-wave symmetry for underdoped compositions to a $s$%
-wave symmetry for the overdoped region \cite{skinta,biswas}. Moreover, it
has been one of the experimental facts that underdoped cuprates possess a
pseudogap \cite{loeser,ding} in the non-superconducting state with anomalous
physical properties below a temperature $T^{\ast }>T_{c}$. The origin of the
pseudogap has become a challenging issue as it might eventually lead to
identify the superconducting mechanism. A class of theoretical models
attempted to describe the pseudogap state as a precursor of the
superconducting $d$-wave gap \cite{emery}.

To date, more experimental results indicate that the charge-ordered stripes
in cuprates represent a specific type of order that has been experimentally
observed to compete with superconductivity \cite%
{kivelson,tranquada,norman,ichikawa}. The stripes are a real-space pattern,
in which the charge carriers condense into rivers of charge separated by
insulating domains \cite{kivelson}. Generally, stripes are classified as
metallic or insulating. It is considered that the static stripe order
(insulating) is bad for superconductivity and is responsible for the
suppression of the superconducting transition temperature \cite{ichikawa},
while the dynamic stripes (metallic) are essential to the superconductivity
in the cuprates \cite{emery,buhler,hotta}. More recently, strong
experimental evidence from the STM studies suggest the presence of a
4-lattice constant checkerboard charge order in underdoped cuprates at the
doping level 1/8, which are referred to as the \textquotedblleft 1/8
anomaly\textquotedblright\ \cite{hanaguri,mcelroy,hoffman,howald,valla}. In
addition, the checkerboard-type ordering of the charge carriers at
\textquotedblleft magic doping fractions\textquotedblright\ have been
intensively studied \cite{komiya,kim,zhou1,zhou2}.\ Castellani \textit{et al.%
} \cite{castellani} have suggested that such commensurate stripes should be
insulating. Furthermore, the underlying crystalline electronic orders (or
Wigner crystals) of charge carriers in cuprates of low doping level have
been proposed \cite{komiya,kim,franz}. In view of these interesting
experimental results and heuristic theoretical proposals, it is now a clear
viewpoint that any interpretation of superconductivity in the cuprates must
account for the real-space charge stripes.

The goal of the present paper is to unveil the mystery of high-$T_{c}$
superconductivity. We argue that the mechanism of superconductivity in both
electron-doped and hole-doped cuprate superconductors should be established
on a picture of pure electrons. We show that this scenario can be used to
explain the basic phenomenology of the high-$T_{c}$ superconductors from the
pairing mechanism to charge stripes. We attempt to provide answers to the
following open questions: Are the pairing and superconducting mechanism of
the electron- and hole-doped cuprates different? What will be the
\textquotedblleft glue \textquotedblright\ to hold the electron pairs
together? What is the one-dimensional nature of the stripes? Are they
insulators or metals? Do the stripes compete with superconductivity, help
superconductivity or hinder it?

\begin{figure}[tbp]
\begin{center}
\resizebox{1\columnwidth}{!}{
\includegraphics{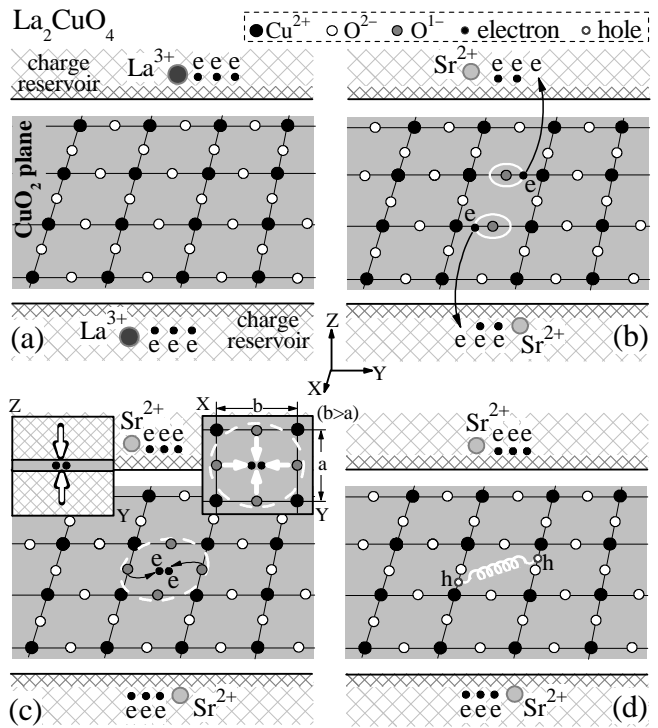}}
\end{center}
\caption{The charge carriers and the Coulomb confinement of electrons in the
CuO plane of cuprate superconductor. (a) The insulating parent compound of La%
$_{2-x}$Sr$_{x}$CuO$_{4}$, (b) the partially substitution of Sr$^{2+}$ for La%
$^{3+}$, (c) two localized charge carriers of undressed electrons, insets
indicate the Coulomb confinement on these electrons, and (d) the
conventional dressed holes picture.}
\label{fig1}
\end{figure}

\section{ Charge carriers: electrons or holes?}

Ever since the discovery of high-temperature superconductivity, many
different cuprate superconductors were synthesized. It should be noted that
though the structures of these materials may be different, the essential
structure to be concerned about is that of the same two-dimensional CuO$_{2}$
planes in them. In particular, the so-called hole-doped La$_{2-x}$Sr$_{x}$CuO%
$_{4}$, due to its relatively simpler structure but a more complex phase
diagram, is the best studied material. The La$_{2}$CuO$_{4}$ [see Fig. \ref%
{fig1}(a)], the parent compound of La$_{2-x}$Sr$_{x}$CuO$_{4},$ is an
insulator. As shown in Fig. \ref{fig1} (b), the partial substitution of Sr$%
^{2+}$ for La$^{3+}$ will result in a deficiency of electrons in the charge
reservoirs and a leaving of electrons from O ions (O$^{2-})$ of CuO$_{2}$
planes to the charge reservoirs. To maintain the local symmetry and the
stability of the CuO$_{2}$ plane, the other two O$^{2-}$ (adjacent the O$%
^{1-}$) tend to donate two electrons which would likely to be trapped in the
center of the four O$^{1-}$ [Fig. \ref{fig1}(c)]. Physically, the effect of
the local distortion pattern in Fig. \ref{fig1}(c) is to minimize the
system's energy. In this case, these two electrons will experience a strong
real space Coulomb confinement [see insets of Fig. \ref{fig1}(c)] imposed by
the charge reservoirs and the local structure distortion in the CuO$_{2}$
plane. It should be noted that the central character (charge carrier) in
this paper is in fact not the artificial hole, but the real electron.

Normally, quantum theory assumes that the charge carriers prefer to stay and
travel (or hopping) in the most \textquotedblleft crowded\textquotedblright\
and complicated Cu$-$O chains, as shown in Fig. \ref{fig1}(d). It now
appears that the paired electrons (or holes) in the CuO$_{2}$ planes has two
possible choices, as shown in Fig. 1(c) and Fig. 1(d). Which situation would
the pair be inclined to choice? If we believe that charge carriers do appear
as suggested by quantum mechanics in the Cu$-$O chains [Fig. \ref{fig1}(d)],
then there will be a strong Coulomb interaction among the charge carriers
and the ions (Cu$^{2+}$ and O$^{2-}$), consequently increase the compound's
energy and this in turn could decrease the system's stability. While for the
case of Fig. \ref{fig1}(c), the Coulomb interaction among the charge
carriers and the ions will be greatly reduced, as a result, increase the
system's stability. Therefore, we argue that the conventional hole-doping
picture of Fig. \ref{fig1}(d) is physically inadequate.

\begin{figure}[tbp]
\begin{center}
\resizebox{0.85\columnwidth}{!}{
\includegraphics{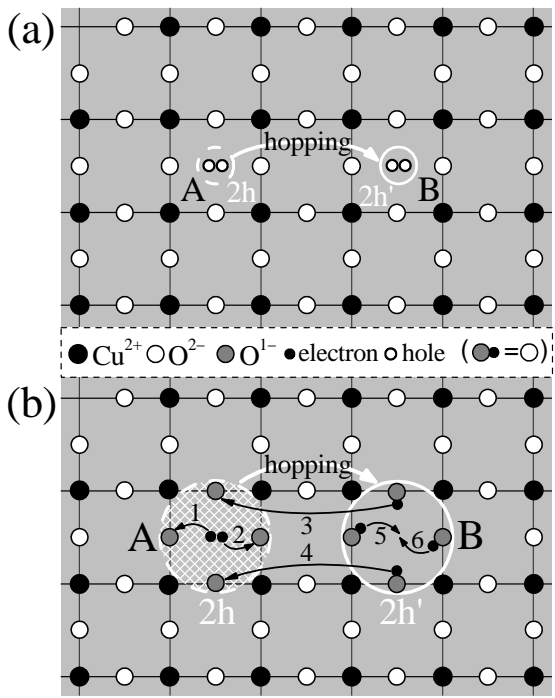}}
\end{center}
\caption{A comparison between an idealized hole pair (sizeless
quasiparticles) and a \textquotedblleft real\textquotedblright\ hole pair.
(a) The idealized hole pair can move (hopping from A to B) easily in CuO$%
_{2} $ (Cu$^{2+}$, O$^{2-})$ plane. (b) The \textquotedblleft
real\textquotedblright\ hole pair can be considered as a dressed electron
pair, the movement of one \textquotedblleft real\textquotedblright\ hole
pair (hopping from A to B) is truly equivalent to the much more complicated
transfer of six electrons (indicated by 1, 2, 3, 4, 5 and 6).}
\label{fig2}
\end{figure}

Having applied the hole concept greatly in solid-state physics, we are faced
with the challenge of constructing an effective physical picture for this
mysterious quasiparticle: What would a hole pair look like if it did exist?
The point of view taken here is that the \textquotedblleft
real\textquotedblright\ holes arise from unique chemical construction in CuO$%
_{2}$ plan. It was assumed that, in the hole doped high-$T_{c}$
superconductors, the hole pairs (the sizeless particles carried a positive
electric charge and a spin) can maintain their integrity in a square lattice
and move [As shown in Fig. \ref{fig2}(a), hopping from A to B] easily and
coherently in CuO$_{2}$ (Cu$^{2+}$, O$^{2-})$ plane \cite{altman}. Here, it
seems worthwhile to point out that this idealized picture of hole is
physically untrue. As shown in Fig. \ref{fig1}(c), when the two electrons
sit on the square lattice, we have a \textquotedblleft
real\textquotedblright\ hole pair pattern as illustrated in Fig. \ref{fig2}%
(b). As one can see from the figure, it appears that the static hole pair
(state A) can be thought of as a dressed electron pair in CuO$_{2}$ (Cu$%
^{2+} $, O$^{2-})$ plane. However, it is shown that this \textquotedblleft
real\textquotedblright\ hole pair is structural unstable. A fatal difficulty
occurs when the pair is moving from A to B [see Fig. \ref{fig2}(b)]. In this
figure, it is clearly shown the simplest possible movement of one
\textquotedblleft real\textquotedblright\ hole pair, in fact, involving a
very complicated transportation of six electrons (there are infinite
possible ways to \textquotedblleft move\textquotedblright\ the pair, note
that only one set of possible movement of the electrons is indicated in the
figure). Therefore, it is hard to imagine that the \textquotedblleft
real\textquotedblright\ hole pair can always maintain its integrity during
the \textquotedblleft hopping\textquotedblright\ process. On this basis, we
argue that the artificial holes should be abandoned.

Anyway, one should be cautious when drawing conclusions based on the concept
of \textquotedblleft hole\textquotedblright . As emphasized by Hirsch, using
the language of `holes' rather than `electrons' in fact obscures the
essential physics since these electrons are the ones that `undress' and
carry the supercurrent (as electrons, not as holes) in the superconducting
state \cite{hirsch}. In the following, we shall illustrate the
superconductive properties of the so-called hole-doped high-$T_{c}$ cuprates
in a pure electron picture (without the hole concept).

\section{ Pairing mechanism, pairing symmetry and pseudogap}

In conventional superconductors, Bardeen, Cooper and Schrieffer (BCS) \cite%
{bcs} argued that phonons (atomic lattice vibrations) act as the
\textquotedblleft glue\textquotedblright\ that binds the electron pairs
together. But at high temperatures, the vibrational motion of the material's
lattice becomes so stiff that it tends to break up the electron pairs
instead of holding them together \cite{anderson0}. So what could possibly
provide the glue that keeps the carriers bound in Cooper pairs? Although
many candidates for this glue (including spin fluctuations, phonons,
polarons, charge stripes and spin stripes) have been proposed. However, what
the pairing glue is in high-$T_{c}$ cuprates is still an open question. In
this Section, I would like to discuss the issue from the point of view of
real-space electron-electron interaction. To test this hypothesis, we are
studying the pairing symmetry of electron- ($n$-type) and hole-doped ($p$%
-type) cuprate superconductors. The section ended with a brief discussion of
pseudogap.

\subsection{ What is the \textquotedblleft glue\textquotedblright\ for
cuprate superconductors?}

\begin{figure}[tbp]
\begin{center}
\resizebox{0.88\columnwidth}{!}{
\includegraphics{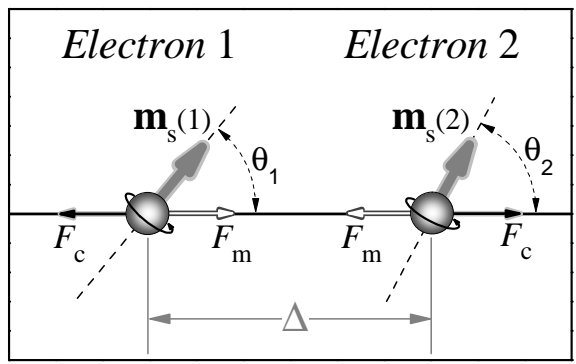}}
\end{center}
\caption{The electromagnetic interaction between two electrons. Normally,
there is a pair of long-range Coulomb repulsion $F_{c}$. In addition, two
spinning electrons may create a pair of short-range attractive forces $F_{m}$
due to the dipolar-dipolar interaction, which can act as the mysterious
\textquotedblleft glue\textquotedblright\ for electron pairs. If \textit{%
electron} 1 and \textit{electron} 2 are free, obviously, the long-range
Coulomb repulsion $F_{c}$ is dominant in comparison with the short-range
attraction $F_{m},$ hence, the $F_{m}$ can work as the \textquotedblleft
glue\textquotedblright\ only if the influence of the repulsion $F_{c}$
between the electrons be successfully suppressed by the external confinement
effects as shown in Fig. \protect\ref{fig1}(c).}
\label{fig3}
\end{figure}

Since the observation of real-space ordering of charge in cuprate
superconductors \cite{kivelson,tranquada,norman,ichikawa}, it is widely
accepted that short-range electron-electron correlations can bind electrons
into real space pairs and dominate the superconductivity properties of the
materials. Normally, as shown in Fig. \ref{fig3}, for the two static
electrons, there is a on-site long-range repulsive electron-electron Coulomb
interaction
\begin{equation}
F_{c}=\frac{e^{2}}{4\pi \varepsilon _{0}\Delta ^{2}},  \label{fc}
\end{equation}%
where $e$ is the electron charge and $\Delta $ is the distance between two
electrons. Due to this strong repulsion of Eq. (\ref{fc}) between electrons,
the confinement scenario of Fig. \ref{fig1}(c) doesn't mean that the
targeted electrons can be naturally paired. So how can repulsive Coulomb
forces exerted on electrons be eliminated so that the electrons can be in
pairs?

It is known that study of superconducting correlations in conventional
superconductors is always performed in momentum-space (dynamic screening), \
where the paired electrons are seldom or never in the same place at the same
time \cite{anderson0}. In the case of dynamic screening, only the long-range
Coulomb interaction $e^{2}/\Delta $ is considered while the short-range
electron--electron magnetic interactions is completely ignored. We argue
here that, in the case of real-space screening, the magnetic forces among
the electrons (see also Fig.\ref{fig3}) could provide a strong
\textquotedblleft glue\textquotedblright\ for electron pairing leading to
high-$T_{c}$ superconductivity in cuprate superconductors. Approximately,
the magnetic dipolar interaction forces $F_{m}$ exerted on the electrons are
given by
\begin{equation}
F_{m}\approx \frac{3\mu _{0}\mu _{B}^{2}}{2\pi \Delta ^{4}}\cos \theta
_{1}\cos \theta _{2},  \label{fm}
\end{equation}%
where $\mu _{0}$ is the permeability of free space and $\mu _{B}$ is the
Bohr magneton. The forces $F_{m}$ can be attractive and repulsive depending
on the orientation ($\theta _{1}$and $\theta _{2}$) of electron magnetic
moment $\mathbf{m}_{s}(j)$, ($j=1,2$). When $\theta _{1}=\theta _{2}=0$ (or $%
\pi )$, the magnetic poles of the paired electrons are lined up in parallel,
consequently, the attractive magnetic force reaches its maximum value $%
F_{m}^{\max }=3\mu _{0}\mu _{B}^{2}/2\pi \Delta ^{4}.$ When $%
F_{c}=F_{m}^{\max },$ we have a stable electron pair with the distance $%
\Delta _{0}$ as

\begin{equation}
\Delta _{0}=\frac{\sqrt{6}\mu _{B}}{ec}\approx 4.73\times 10^{-3}\mathring{A}%
,  \label{distant}
\end{equation}%
where $c$ is the speed of light in vacuum.

\begin{figure}[tbp]
\begin{center}
\resizebox{0.8\columnwidth}{!}{
\includegraphics{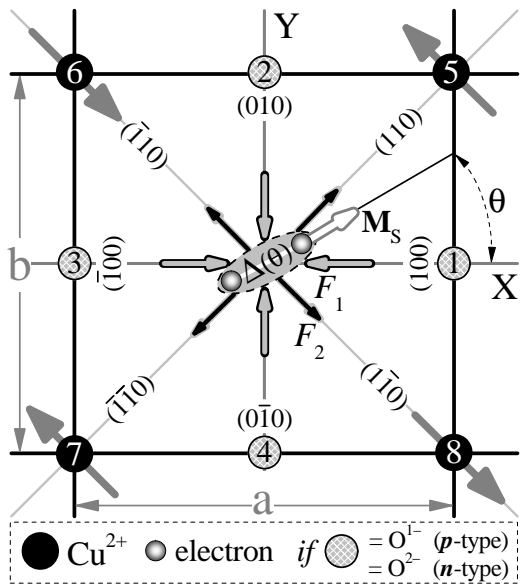}}
\end{center}
\caption{Two spin parallel electrons with a joint magnetic moment $\mathbf{M}%
_{s}$ is confined inside one unit cell of CuO plane. Only nearest-neighbor
(1, 2, 3, 4) and next-nearest-neighbor (5, 6, 7, 8) interactions are
considered. If $\Delta (\protect\theta )\ll a$, approximately, there are
four confinement forces ($F_{1}$) in both the $x$ and $y$ directions, while
four dissociated forces ($F_{2}$) in the diagonal directions.}
\label{fig4}
\end{figure}

Because of the short-range interaction characteristics of Eq. (\ref{fm}), as
is usually the case $F_{c}\gg F_{m}^{\max }$, then it is clear that the
\textquotedblleft glue\textquotedblright\ (attractive magnetic force $%
F_{m}^{\max }$) works only if the cuprate superconductors has the effect of
weakening the long-range repulsive force $F_{c}$. We presume that the
real-space confinement effect (electromagnetic interactions) in CuO$_{2}$
plane [see Fig. \ref{fig1}(c)] plays a central role in suppressing the
influence of the Coulomb repulsion between electrons. To describe this, two
spin parallel electrons of Fig. \ref{fig3} with a joint paired-electron
magnetic moment $\mathbf{M}_{s}=\mathbf{m}_{s}(1)+\mathbf{m}_{s}(2)$ are
embedded into a CuO plane of the cuprate superconductor, as shown in Fig. %
\ref{fig4}. Looking at the figure, just a simplification, only
nearest-neighbor and next-nearest-neighbor interactions are considered.\
Inside the unit cell, the possible paired-electrons with the magnetic moment
$\mathbf{M}_{s}$ along the $\theta $ direction, and the corresponding
distance between the electrons is reexpressed as $\Delta (\theta )$. From
the figure, one can easily conclude that the pair with the $\mathbf{M}_{s}$
oriented in (100), (010), ($\overline{1}$00) and (0$\overline{1}$0)
directions is generally considered to be much more stable due to the
suppression ($F_{1}$) of the four oxygen ions (O$^{1-}$ for hole-doped, O$%
^{2-}$ for electron-doped), as opposed to the cases, in (110), ($\overline{1}
$10), ($\overline{1}\overline{1}$0) and (1$\overline{1}$0) directions where
the bound pair tends to be separated by Coulomb forces ($F_{2}$) of the Cu$%
^{2+}$. As a consequence, the distance $\Delta (\theta )$ between the two
electrons of the pair has a minimum at $\theta =0$, $\pi /2$, $\pi $ and 3$%
\pi /2$, while at $\theta =\pi /4$, $3\pi /4$, $5\pi /4$ and $7\pi /4$, $%
\Delta (\theta )$ will reach its maximum value. According to Eq. (\ref{fm}),
the orientation dependence of binding energy for the electron pair within a
single plaquette can be defined as

\begin{equation}
E_{b}(\theta )=\frac{\mu _{0}\mu _{B}^{2}}{2\pi \Delta ^{3}(\theta )}\approx
\frac{1.074\times 10^{-3}}{\Delta ^{3}(\theta )}(eV),  \label{binding}
\end{equation}%
where the unit of distance $\Delta (\theta )$ is angstrom.

It is important to note that no quasiparticles are involved in the above
studies, hence the presented real space pairing mechanism (\textquotedblleft
glue\textquotedblright ) can be regarded as a self-pairing mechanism due to
the intrinsic electromagnetic exchange coupling between the real electrons.
There is no reason to believe that the dynamic screening (quasiparticle
approximation) is related to a \textquotedblleft glue\textquotedblright\ in
the high-$T_{c}$ cuprates where the net interaction contributed by
\textquotedblleft quasiparticle\textquotedblright\ is found always
repulsive, as emphasized by Anderson \cite{anderson0}.

\subsection{Pairing symmetry in $n$-type and $p$-type cuprate
superconductors.}

Now we turn to the pairing symmetry. In recent years a number of experiments
have indicated that the predominant real-space $d$-wave symmetry in
hole-doped cuprate superconductors \cite{harlingen}. Theoretical and
numerical studies of the two-dimensional Hubbard $t$-$J$ models have also
suggested $d$-wave pairing symmetry \cite{white}. Altman and Auerbach have
provided an intuitive picture of how $d$-wave hole pair can survive in a
single plaquette \cite{altman}. On the other hand, pairing symmetry in
electron-doped cuprate superconductors, such as Nd$_{2-x}$Ce$_{x}$CuO$_{4}$
(NCCO) \cite{tokura}, is still a controversial topic \cite%
{tsuei,ariando,skinta,biswas,tsuei2,qazilbash}.

Here, we argue that pairing symmetry in cuprate superconductors may also be
explained analytically in the picture of the real space Coulomb confinement
effect of Fig. \ref{fig4}. In the case of hole-doped copper oxides, as shown
in Fig. \ref{fig4}, the pair with $\mathbf{M}_{s}$ parallel to $x$- and $y$%
-axis has a minimum $\Delta (\theta )$ that leads to a maximum binding
energy of Eq. (\ref{binding}), namely, the maximum energy gap in these four
directions. Furthermore, it is possible that the pair ($\mathbf{M}_{s}$
along the diagonal directions) can be destroyed $\left[ \Delta (\theta )>%
\sqrt{2}a\right] $ due to the strong Coulombic interaction between the pair
and ions Cu$^{2+},$ hence, the corresponding binding energies may be zero in
these four directions. Within the proposed real-space confinement mechanism,
the pairing picture for the electron-doped systems is remarkably similar to
that of the hole-doped cuprates, with one major difference comes from the
substitution of O$^{1-}$ by O$^{2-}$ (see also Fig. \ref{fig4}). This
substitution, on the one hand, greatly enhance the confinement effect by
increasing $F_{1}$ in (100), (010), ($\overline{1}$00) and (0$\overline{1}$%
0) directions and in turn will shorten the distance $\Delta (\theta )$ and
increase the binding energy, on the other hand, it can greatly reduce the
possibility of the \textquotedblleft pair-breaking\textquotedblright\ along
directions (110), ($\overline{1}$10), ($\overline{1}\overline{1}$0) and (1$%
\overline{1}$0), in other words, now the diagonal's pair is more likely to
survive in the electron-doped cuprates than in the hole-doped cuprates.
Roughly speaking, our unified model for cuprate superconductivity yields for
the hole-doped (\textit{p}-type) and electron-doped (\textit{n}-type)
cuprates to have essentially the same pairing mechanisms, but different
spatial pairing symmetries.
\begin{figure}[tbp]
\begin{center}
\resizebox{0.9\columnwidth}{!}{
\includegraphics{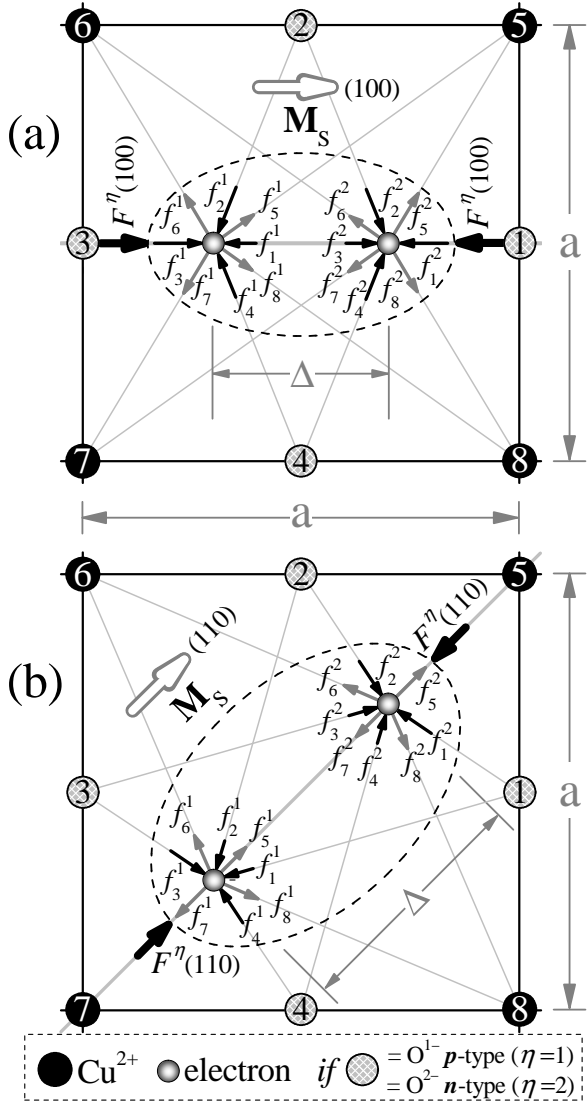}}
\end{center}
\caption{The schematic plot of the confinement forces acting the electron
pair inside one unit cell of the CuO plane. The pair oriented ($\mathbf{M}%
_{s}$) in (a) (100) direction ($\protect\theta =0)$, and (b) (110) direction
($\protect\theta =\protect\pi /4)$. }
\label{fig5}
\end{figure}

To make the arguments of pairing symmetry more convincing, below more
detailed studies will be done based on the simple Coulomb's equation. As
shown in Fig. \ref{fig4}, in the case of the nearest-neighbor (marked by 1,
2, 3, 4) and next-nearest-neighbor interactions (marked by 5, 6, 7, 8), the
forces acting on (push/pull) the pair are in eight different directions.
Within this framework, the pair will stay at the most unstable state in the
pair's orientation due to the strong repulsive Coulomb interaction between
the two electrons in this direction, we are therefore paying particular
attention to the confinement effect in the same direction as the orientation
of the pair. If the electron pair is oriented in $x$- and $y$-axis or the
diagonal directions, because of the structural symmetry, we can present the
explicit analytical expressions of the confinement effects. Figures \ref%
{fig5}(a) and (b) illustrate the eight Coulombic forces and the total
confinement forces [$F^{\eta }(100),$ $F^{\eta }(110)$] exerted on the two
electrons of the pair in (100) and (110) directions of the cuprate samples,
respectively. It is easy to have some similar figures in other directions
for both $p$-type and $n$-type superconductors. According to Fig. \ref{fig5}%
, we can get a general formula of the total confinement force $F^{\eta
}(100) $ applied to the electron of the pair in (100) direction as%
\begin{equation}
F^{\eta }(100)=\frac{e^{2}}{4\pi \varepsilon _{0}}\left[ \eta \left( \frac{1%
}{d_{1}^{2}}-\frac{1}{d_{2}^{2}}\right) +\frac{1}{d_{3}^{2}}-\frac{1}{%
d_{4}^{2}}\right] ,  \label{force100}
\end{equation}%
where $\eta =1$ for hole-doped case, while $\eta =2$ for electron-doped
cuprates. And the parameters $d_{1}$, $d_{2},$ $d_{3}$ and $d_{4}$ are
defined by
\begin{eqnarray*}
d_{1} &=&\frac{\sqrt{a^{2}-\Delta ^{2}}}{4\sqrt{a\Delta }},\quad d_{2}=\frac{%
\left( a^{2}+\Delta ^{2}\right) ^{3/4}}{2\sqrt{2\Delta }}, \\
d_{3} &=&\frac{(2a^{2}+\Delta ^{2}+2a\Delta )^{1/4}\sqrt{2a^{2}+\Delta ^{2}+%
\sqrt{2}a\Delta }}{4\sqrt{a+\Delta }}, \\
d_{4} &=&\frac{(2a^{2}+\Delta ^{2}-2a\Delta )^{1/4}\sqrt{2a^{2}+\Delta ^{2}-%
\sqrt{2}a\Delta }}{4\sqrt{a-\Delta }},
\end{eqnarray*}%
where $\Delta <a.$ Similarly, in (110) direction, we have

\begin{equation}
F^{\eta }(110)=\frac{e^{2}}{4\pi \varepsilon _{0}}\left[ \frac{1}{D_{1}^{2}}-%
\frac{1}{D_{2}^{2}}+\eta \left( \frac{1}{D_{3}^{2}}-\frac{1}{D_{4}^{2}}%
\right) \right] .  \label{force110}
\end{equation}%
Here too $\eta =1$ and $\eta =2$ for hole- and electron-doped cuprates,
respectively. The four distance parameters in Eq. (\ref{force110}) are given
by
\begin{eqnarray*}
D_{1} &=&\frac{\left( 2a^{2}+\Delta ^{2}\right) ^{3/4}}{4\sqrt{\Delta }}%
,\quad D_{2}=\frac{(2a^{2}-\Delta ^{2})}{4\times 2^{3/4}\sqrt{a\Delta }}, \\
D_{3} &=&\frac{(a^{2}+\Delta ^{2}-\sqrt{2}a\Delta )^{3/4}}{2\sqrt{\sqrt{2}%
a-2\Delta }}, \\
D_{4} &=&\frac{(a^{2}+\Delta ^{2}+\sqrt{2}a\Delta )^{3/4}}{2\sqrt{\sqrt{2}%
a+2\Delta }},
\end{eqnarray*}%
where $\Delta <a/\sqrt{2}.$
\begin{figure}[tbp]
\begin{center}
\resizebox{0.85\columnwidth}{!}{
\includegraphics{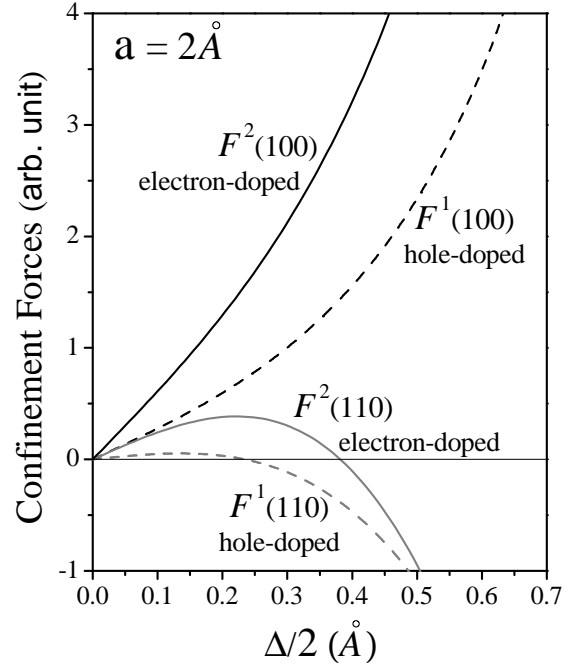}}
\end{center}
\caption{Analytical confinement forces versus $\Delta /2$ in electron- and
hole-doped cuprate superconductors along (110) and (100) directions.\ In the
case of electron doping (solid lines), when $\Delta $ is small, both
confinement forces in (110) and (100) are positive, but note that (100)
confinement (black solid line) is stronger than (110) confinement (gray
solid line), these yield the ($s+d$) symmetry in the system. In the case of
hole doping (dashed lines), for a small $\Delta $, there is a strong
confinement in (100) direction (black dashed line), while a very weak
confinement in (110) direction (gray dashed line), these results indicate a
dominating $d$-wave pairing symmetry (with a small $s$-wave component) in
hole-doped cuprates.}
\label{fig6}
\end{figure}

\begin{figure}[tbp]
\begin{center}
\resizebox{1\columnwidth}{!}{
\includegraphics{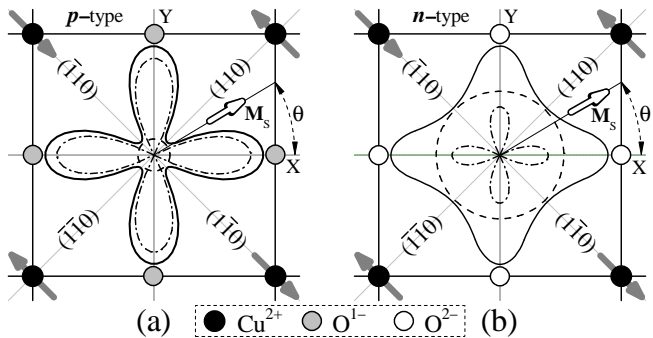}}
\end{center}
\caption{Pairing symmetry in cuprate superconductors, (a) hole-doped, (b)
electron-doped. The pairing electrons maintain their integrity inside the
single plaquette with the dominant $d$-wave (a mixture of large $d$-wave and
small $s$-wave components) and $(s+d)$-wave symmetry in $p$-type and $n$%
-type samples, respectively.}
\label{fig7}
\end{figure}

From equations (\ref{force100}) and (\ref{force110}), if $F^{\eta }(100)>0,$
the total confinement is \textquotedblleft positive\textquotedblright\ and
the pair tends to be stable in (100)-($\overline{1}$00) directions, while $%
F^{\eta }(100)<0,$ the total confinement is \textquotedblleft
negative\textquotedblright\ and the pair with an apparent tendency to be
broken up in these directions. Thus the situation is the same as for the $%
F^{\eta }(110).$ With the analytical expressions (\ref{force100}) and (\ref%
{force110}), we draw in Fig. \ref{fig6} the confinement forces versus $%
\Delta /2$ (two solid lines for electron-doped case, while the dash lines
for hole-doped sample) under the condition $a=2\mathring{A}$. This figure
reveals three important facts: (i) in the case of hole doping, the existence
of a strong confinement in (100) direction as expected, a surprising result
is that, there is a very weak confinement in (110) direction; (ii) in the
case of electron doping, if $\Delta /2\leq 0.38\mathring{A}$, both
confinement forces in (110) and (100) are positive, but note that (100)
confinement is stronger than (110) confinement, (iii) in both cases (hole
and electron doping), the net confinement $F^{\eta }(100)$ is always
positive inside the unit cell. Having obtained the strongest confinement $%
F^{\eta }(100)$ and the weakest confinement $F^{\eta }(110)$ for both hole-
and electron-doped cuprates (see Fig. \ref{fig6}), now we can qualitatively
determine the pairing symmetries for two different doped systems, as shown
in Fig. \ref{fig7}. The solid black curve of Fig. \ref{fig7}(a) shows a
mixture of large $d$-wave (dash-dotted curve) and small $s$-wave (dashed
circle) pairing symmetries in hole-doped cuprates. Our theoretical result of
Fig. \ref{fig7}(a) is in satisfactory agreement with the experiments
indicating the existence of a very small $s$-wave component in hole-doped
cuprates \cite{kouznetsov,li} Also, basing on the Fig. \ref{fig6}, the
results definitely favor the ($s+d$) symmetry in electron-doped cuprates, as
shown in Fig. \ref{fig7}(b).

In fact, the above discussions have been restricted to the idealized
condition of zero temperature. If temperature induced thermal effects
(lattice vibration) are taken into account, the obvious difference [Fig. \ref%
{fig7}(a) and (b)] in pairing symmetry between hole- and electron-doped
cuprates can be further interpreted from the viewpoint of the
superconducting transition temperature $T_{c}$. It is well known that the
maximum $T_{c}$\ in the hole-doped cuprates is 133 $K$, which is much higher
than that of the electron-doped systems (about 30 $K$), for this reason, a
larger dynamical local lattice distortion in the CuO plane can be expected
in the hole-doped cuprates. As a result, the electron pair is more likely to
be broken up by ions Cu$^{2+}$ along the diagonal directions of Fig. \ref%
{fig7}(a), this discussion provides a further support the predominantly $d$%
-wave superconducting symmetry in the hole-doped cuprates. Lowering the
cuprates temperature will possibly result in a symmetry transition from $d$%
-wave to ($s+d$)-wave in hole-doped systems. We think the suggested scenario
offers a new way for physical interpretation of the pairing symmetry in the
cuprates.

\subsection{Pseudogap in the cuprate superconductors}

The nature of the normal-state gap (pseudogap) phase of HTSC is also highly
controversial \cite{armitage1,basov,deutscher,renner,loeser1,ding2,marshall}%
. ARPES and tunneling measurements show a clear pseudogap which was seen to
persist even at room temperature \cite{loeser,renner,ding2}. There are many
models attempt to describe the mysterious pseudogap state. Strictly
speaking, none of the proposed models is completely satisfactory. As
discussion above, here we present a new approach based on the simple and
natural picture of the real-space confinement effect, and the pseudogap is
associated with the local structure of unit cell in CuO$_{2}$ plane. Thus it
should not be surprising about the pseudogap behavior which indicate the
formation of pairs below $T^{\ast }>T_{c}.$ The real space Coulomb
confinement picture (see Fig. \ref{fig4}) shows clearly that the pair is
confined within one single plaquette, consequently, the \textquotedblleft
size\textquotedblright\ of the electron pair [distance $\Delta (\theta )$]
is much less than one lattice constant. For $\Delta (\theta )<0.22\mathring{A%
},$ the binding energy of Eq. (\ref{binding}) may be much larger than the
magnetic Heisenberg exchange ($J\sim 0.1$eV). This support the experiments
that the pseudogap can exist in the cuprate superconductors at a temperature
even higher than room temperature.

The presence of pseudogap in underdoped cuprate superconductors has been a
solid experimental fact. However, whether it is a continuation of the
superconducting gap as a result of pre-formed pairs is a topic of current
debate. Our results suggest that the pseudogap is a general feature of all
transition metal oxides, not just cuprate superconductors. The newly
observations of the\ Cooper pair localization in the amorphous insulating
systems \cite{stewart}, in fact, reveal that pair correlations may be built
up in the real space confinement screening and the pseudogap is a common
feature in nature materials that is consistent with our arguments in this
section.

\section{Crystalline electron pairs, charge stripes and spin density wave}

Physically, pairing in cuprates is an individual behavior characterized by
pseudogap, while superconductivity is a collective behavior of many coherent
electron pairs. Nowadays, more and more beautiful experimental results
suggest that stripes are common in cuprates and may be important in the
mechanism for HTSC \cite%
{hanaguri,tranquada,norman,kim,noda,zhou,mook,vershinin,wakimoto,keimer,kastner,momono}%
. Here, we argue that there are only two specific stripes which would
contribute to the mechanism of superconductivity in cuprate superconductors
and the dynamical spin density wave (SDW) coherent phases can be established
along the stripes.

\subsection{Wigner crystal and charge stripes}

\begin{figure}[tbp]
\begin{center}
\resizebox{1\columnwidth}{!}{
\includegraphics{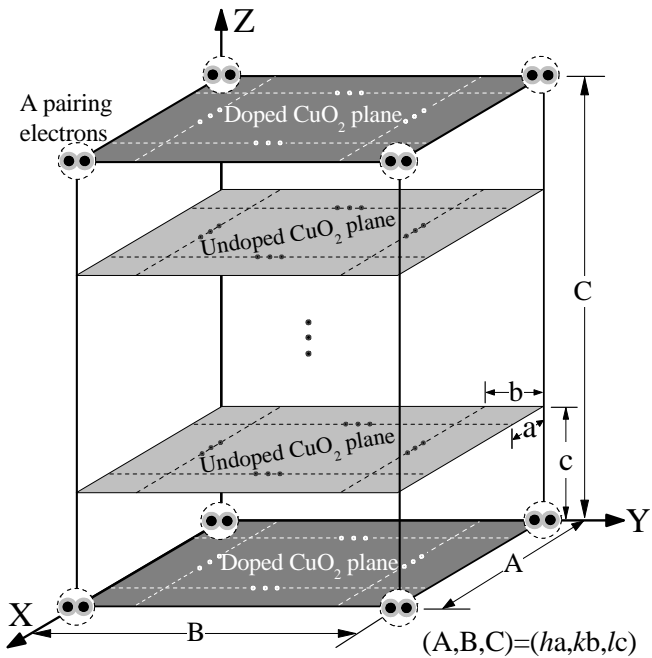}}
\end{center}
\caption{Simplified schematic unitcell of the electron-pairs (dimerized)
Wigner crystal in the high-$T_{c}$ cuprates.}
\label{fig8}
\end{figure}

In nature, periodic structures are often considered as the result of
competition between different interactions. The formation of stripe patterns
is generally attributed to the competition between short-range attractive
forces and long-range repulsive forces \cite{seul}. Obviously, our scenario
provides not only the \textquotedblleft glue\textquotedblright\ for the
electron pairs but also the most basic competitive environment [attraction $%
F_{m}$ of Eq. (\ref{fm}) and repulsion $F_{e}$ of Eq. (\ref{fc}) among the
electron pairs] for the possible formation of charge stripe. In this paper,
we focus on the doping-induced behaviors of La$_{2-x}$Sr$_{x}$CuO$_{4}$with
the primitive cell ($a,b,c$). At a rather low doping level, the interactions
among electron pairs can be neglected and the superconductor behaves much
like a charged random system. However, as more carriers are added, the
effect of the competitive interactions among electron pairs will emerge. As
a result, at a proper doping level the electron pairs can self-organize into
a `superlattice' (Wigner crystal of electron pairs) with the primitive cell $%
(A,B,C)=(ha,kb,lc)$, as shown in Fig. \ref{fig8}. Consequently, the
\textquotedblleft material\textquotedblright\ composed of electron-pair
\textquotedblleft atoms\textquotedblright\ will undergo a structure
transition from random to order phase (LTO or LTT). Thus, the carrier
density $x$ is given by

\begin{equation}
x=p(h,k,l)=2\times \frac{1}{h}\times \frac{1}{k}\times \frac{1}{l},
\label{fractions}
\end{equation}%
where $h$, $k$, and $l$ are integral numbers. As can be seen, the
`superlattice' scheme is most consistent with the recent experimental
observations of the two-dimensional nature of the charge stripes \cite%
{kim,christensen,hayden}. Expression (\ref{fractions}) plays a key role in
the following studies.

\begin{figure}[tbp]
\begin{center}
\resizebox{1\columnwidth}{!}{
\includegraphics{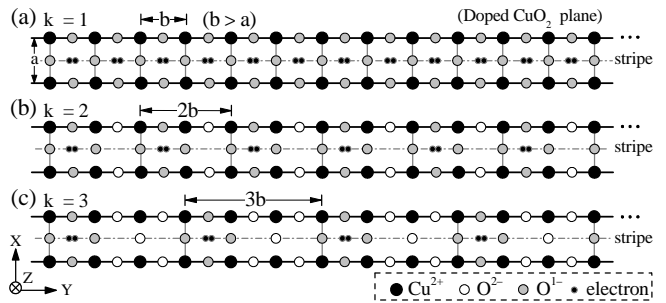}}
\end{center}
\caption{The quasi-one-dimensional charge stripes in high-T$_{c}$ cuprates,
(a) and (b) are the metallic charge stripes, where the central chains are
composed of mono-ion (O$^{1-})$, while (c), the insulating charge stripe,
there are two different ions (Cu$^{2+}$ and O$^{1-})$ in the central chain.}
\label{fig9}
\end{figure}

When the charge stripe spacing satisfying $A\gg B,$ the
quasi-one-dimensional striped phase can be expected \cite{noda,zhou,mook}.
In this case, we can devote most of our attention to the ladder structures
of Fig. \ref{fig9}. One particularly important conclusion can be drawn from
this figure. When $k=1$ or $2$, inside the ladder [Fig. \ref{fig9}(a) and
(b)] there is an perfect mono-ion (O$^{1-})$ chain where the coherence can
be easily established among electron pairs, hence the corresponding charge
stripes are metallic. In fact, the new type of orders of Fig. \ref{fig9}(a)
and (b) show the typical case of a charge-Peierls dimerized transition. As
shown in Fig. \ref{fig9}(c), while for $k\geq 3$, the middle chain is now
impure (containing two kinds of oxygen ions O$^{1-}$ and O$^{2-}$) and this
will greatly restrain the establishment of coherence among electron pairs,
as a result the strips will behave more like the one-dimensional insulators
\cite{noda,zhou}. It appears, then, that the pure and perfect oxygen ion (O$%
^{1-})$ chains inside the charge stripes play a central role in the
superconductivity. In addition, the one-dimensional nature (the charge
stripes extend along $b$) indicates that the Coulombic interactions between
two nearest neighbor electron pairs alone the $b$ direction is much stronger
than that of electron pairs along $a$, then the anisotropic lattices ($b>a$)
should be found experimentally in the CuO$_{2}$ plane \cite{bednorz,wumk}.

\subsection{Spin density wave (SDW)}

Currently, the magnetic (spin) excitation has been observed by neutron
scattering in a number of hole-doped \cite{dai,sternlieb,arai,fauque} and
electron-doped materials \cite{wilson,ismer}. It is generally believed that
magnetic excitations might play a fundamental role in the superconducting
mechanism. Our results above demonstrate that the short-range real space
electron-electron interactions are relevant to the pairing symmetry
(pseudogap) and we would expect in the following that the magnetic
interactions (magnetic resonance) among electron pairs is essential in the
mechanism of superconductivity. It should be noted that Fig. \ref{fig9} is
merely a schematic picture which is by no means a complete restriction of
the stripes in the center of ladders. Evidently, a direct interaction among
electron pairs and the domain-walls (Cu$^{2+}-$O$^{1-}$ chains for $k=1$)
will result in the fluctuations of stripes which have been confirmed by
experiments \cite{noda,zhou,mook,wakimoto}.

It is well known that the concept of the superconducting order parameter,
which was first introduced by Ginzburg and Landau (GL) \cite{ginzburg} in
their study of the superconducting state in 1950, is a very important factor
for the qualitatively description of the ordered state of various phase
transitions. To describe the competition between superconductivity and
charge stripe order in Fig. \ref{fig9}, based on the GL theory formalism, a
complex position-dependent order parameter can be defined as

\begin{figure}[tbp]
\begin{center}
\resizebox{0.95\columnwidth}{!}{
\includegraphics{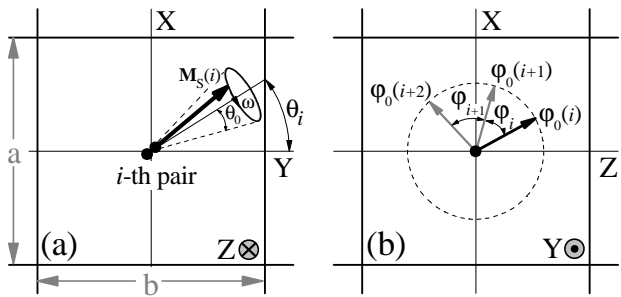}}
\end{center}
\caption{The initial magnetic phase relationship of the electron pairs in
the charge stripe of Fig. \protect\ref{fig9}(a). (a) The $i$-th electron
pair with $\mathbf{M}_{s}(i)$ is oriented at angle $\protect\theta _{i}$ and
$\protect\theta _{0}$ corresponds to a small-angle precession (fluctuations)
due to the thermal vibrations of the lattices, (b) the initial magnetic
phase difference between two nearest-neighbor pairs. In the
non-superconductivity state, $\protect\theta _{i}$ and $\protect\varphi _{i}$
are random; while in superconductivity state, $\protect\theta _{i}$ and $%
\protect\varphi _{i}$ will correspond to a constant, respectively.}
\label{fig10}
\end{figure}
\begin{figure}[tbp]
\begin{center}
\resizebox{1\columnwidth}{!}{
\includegraphics{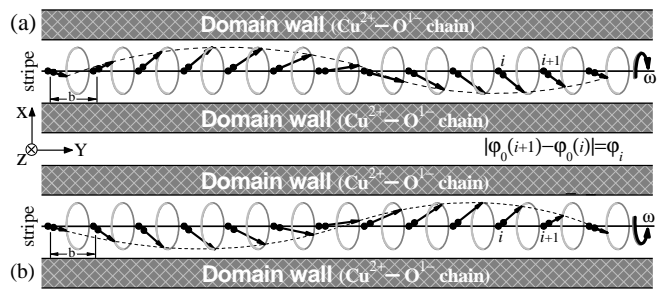}}
\end{center}
\caption{Due to the magnetic phase-coherence (the initial phase difference $%
\protect\varphi _{i}=\protect\varphi _{0}$ is a constant) among the electron
pairs, a helical dynamical spin-density-wave (DSDW) is inspired in the
metallic charge stripe of Fig. \protect\ref{fig9}(a) and the
superconductivity and DSDW can coexist along this stripe, (a) The left-hand
SDW, and (b) the right-hand SDW. Similar dynamical SDW can also exist in
Fig. \protect\ref{fig9}(b).}
\label{fig11}
\end{figure}


\begin{equation*}
\Psi =\Lambda e^{i\Phi },
\end{equation*}%
and parameters $\Lambda $ and $\Phi $ are
\begin{eqnarray*}
\Lambda &=&\frac{1}{n}\sum\limits_{i}\mathbf{M}_{s}(i), \\
\Phi &=&\frac{1}{n}\sum\limits_{i}\left[ \varphi _{0}(i+1)-\varphi _{0}(i)%
\right] =\frac{1}{n}\sum\limits_{i}\varphi _{i},
\end{eqnarray*}%
where $\mathbf{M}_{s}(i)$ is the joint magnetic moment of the $i$-th
electron pair, and $\varphi _{0}(i)$ and $\varphi _{i}$ are the initial
phase and the initial phase difference between two nearest-neighbor pairs
(see Fig. \ref{fig10}), respectively. As shown in Fig. \ref{fig10}(a), the $%
\mathbf{M}_{s}(i)$ is oriented at angle $\theta _{i}$ and $\theta _{0}$
corresponds to a small-angle precession due to the following two reasons:
(i) a short-range magnetic interaction between the $\mathbf{M}_{s}(i)$ and
antiferromagnetic background of CuO plane; (ii) the thermal vibrations of
the lattices. Fig. \ref{fig10}(b) shows the initial phases and the initial
phase differences of the phase-incoherent electron pairs inside the charge
stripe of Fig. \ref{fig9}(a). Without an external field, $\theta _{i}$ and $%
\varphi _{i}$ are random, hence $\Lambda $, $\Phi $ and the order parameter $%
\Psi $ are zero. In such a situation, although all electrons may have been
paired under appropriate temperature $T^{\ast }$, they are completely
incoherent and the superconductivity might not be expected in the charge
stripes. When $T<T_{c}$ and an external field is applied (for example, in $y$%
-direction) to the cuprate sample, in this case $\theta _{i}=0,$ $\theta
_{0}\neq 0$ and $\varphi _{i}=\varphi _{0}$ is a constant, then $\Lambda $, $%
\Phi $ and $\Psi $\ reach their maximum values, indicating the occurrence of
symmetry breaking and the superconductive transition. As a result, a real
space helical dynamical spin-density-wave (SDW) (segregated by the domain
wall) and superconductivity coexist to form a supersolid, as shown in Fig. %
\ref{fig11}. Therefore, if stripes are to have a positive relevance to
superconductivity in the cuprates, then they must be able to exist in a
dynamic magnetic coherent form. It is obvious that a similar superconductive
stripe phase can occur in Fig. \ref{fig9}(b). Now, the real space helical
SDW is proved to be a much richer natural phenomenon \cite{uchida}, not
restricted to cuprate superconductors.

\section{Magic doping fractions}

\label{magic}

In fact, one particularly important characteristic of the HTSC is the
competition among different order phases. This competition could emerge as a
result of some intriguing features such as the \textquotedblleft magic
doping fractions\textquotedblright\ in high-$T_{c}$ superconductors \cite%
{komiya}. It is of much interest to note that our mechanism offers a natural
explanation of the physical origin of these observations \cite%
{kim,noda,zhou,mook,christensen,hayden}. Generally, the \textquotedblleft
superlattice\textquotedblright\ (electron pairs) exist in LTO phases. In
this paper, it is argued that the \textquotedblleft magic doping
phases\textquotedblright\ in cuprates can exist in the LTT \textquotedblleft
superlattice\textquotedblright\ phases. Moreover, it is shown that there are
two different kinds of LTT \textquotedblleft superlattice\textquotedblright\
phases: LTT1 phases (where $A=B\neq C$ and $a=b$), in which the square
superstructure of electron pairs is inside the CuO$_{2}$ plane; LTT2 phases
(where $B\ll A=C$ and $b>a$), in which the square superstructure is
perpendicular to the CuO$_{2}$ plane. Further studies indicate that the LTT1
phases are related to the anomalous suppression of superconductivity, while
the LTT2 phases would contribute to the mechanism of superconductivity in
cuprate superconductors.

\subsection{Underdoped $(x=1/18$, 1/8, 1/16, 1/9$)$}

\begin{figure}[tbp]
\begin{center}
\resizebox{1\columnwidth}{!}{
\includegraphics{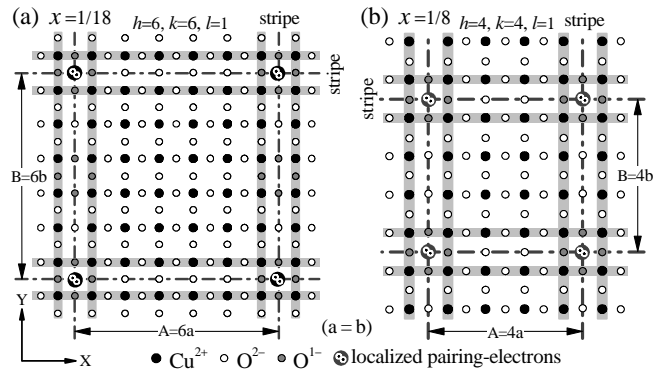}}
\end{center}
\caption{The nondispersive superlattices of the electron pairs in the doped
CuO$_{2}$ planes. (a) $x=1/18$, and (b) $x=1/8$. Both the superlattice of
electron pairs and the original lattice are in the LTT phase, especially the
superstructures with a commensurate periodicity of lattice constants are
formed within the Cu-O planes.}
\label{fig12}
\end{figure}

The experimental and theoretical results demonstrate that the
insulator-to-superconductor transition in the underdoped regime ($x\approx
0.055$) in LSCO \cite{wakimoto,keimer,kastner,sugai}. We note that the
observed decimal number ($0.055$) is coincident with a rational doping level
($\sim 1/18$). At $x=1/18,$ there are some possible configurations, [for
instance, $p(6,6,1)$, $p(6,3,2)$, $p(12,1,3)$, $p(18,1,2)$ $\cdots ],$ which
satisfy Eq. (\ref{fractions}). From the viewpoint of the stripe stability,
the charge orders tend to choose the LTT1 phase of $p(6,6,1)$ and the
corresponding doped CuO$_{2}$ plane is shown in Fig. \ref{fig12}(a). It can
be seen from this figure that all the paired electrons are localized at the
lattice points of the superlattice and thereby hinder the superconductivity.
At $x=1/8,$ another LTT1 \textquotedblleft superlattice\textquotedblright\
phase of $p(4,4,1)$ can also coexists with the LTT original lattice ($a=b$)
of the LSCO [see Fig. \ref{fig12}(b)]. This may explain the famous
\textquotedblleft 1/8 anomaly\textquotedblright\ in various high-$T_{c}$
superconductors \cite{tranquada,valla,moodenbaugh,crawford,homes,fujita}. At
the doping level 1/8, results of recent STM experiments on Bi$_{2}$Sr$_{2}$%
CaCu$_{2}$O$_{x}$ (BSCCO) \cite{vershinin} and Ca$_{2-x}$Na$_{x}$CuO$_{2}$Cl$%
_{2}$ (CNCOC) \cite{hanaguri} suggest a checkerboard-like spatial modulation
of electronic density of states with a periodicity of $4a\times 4a$. Several
theoretical scenarios have been proposed to explain the commensuration
charge ordering patterns \cite{hanaguri,komiya,altman,melikyan,wrobel}. We
consider that the schematic [Fig. \ref{fig8} and Eq. (\ref{fractions})] is a
promising approach to the problem, as it can explain the longstanding puzzle
in a very natural way.

Specifically, we find that this simple picture can provide some preliminary
evidence for the other anomalous behaviors in cuprates. From Eq. (\ref%
{fractions}), it is clear that the nondispersive superlattices of $4a\times
4a$ and $3a\times 3a$ in CuO$_{2}$ planes can be expected at $x=1/16$ (see
Kim \textit{et al.} for details) \cite{kim} of $p(4,4,2)$ and $x=1/9$ of $%
p(3,3,2)$, respectively. Therefore, the anomalous suppression of the
superconductivity can be found in LSCO at $x=1/9$ and $x=1/16$.
Encouragingly, some unusual results have already been observed at $x=1/16$
and $x=1/9$ of the underdoped LSCO crystals. For instance, by high
resolution ARPES experiments on $x\sim 1/16$ sample, an anomalous change at $%
\sim 70$ mev in the nodal scattering rate was reported \cite{zhou1}, and the
observations of intrinsic anomalous superconducting properties at magic
doping levels of $x=1/16$ and $x=1/9$ had been found by $dc$ magnetic
measurements \cite{zhou2}. It is important to mention that the present
results well explain an open question: why two different LSCO compounds ($%
x=1/8$ and $x=1/16$) can exhibit the same nondispersive $4a\times 4a$
superstructure within their CuO$_{2}$ planes.

Neutron scattering provides a direct measure of the magnetic excitation
spectrum. High resolution neutron scattering experiments on optimally doped
La$_{2-x}$Sr$_{x}$CuO$_{4}$ ($x=0.16$) \cite{christensen} and YBa$_{2}$Cu$%
_{3}$O$_{6+x}$ \cite{daipc} reveal that the magnetic excitations are
dispersive. However, a similar study on La$_{2-x}$Ba$_{x}$CuO$_{4}$ ($x=1/8$%
) shows some nondispersive superlattice peaks indicative of spin and charge
stripe order \cite{fujita}. In particular, in La$_{1.6-x}$Nd$_{0.4}$Sr$_{x}$%
CuO$_{4}$ for $x=1/8$, the static SDW and charge-density-wave (CDW) have
been discovered \cite{tranquada}. The same observations have been reported
in other cuprates \cite{wakimoto,fujita,mitrovic}. Note that the static SDW
and static CDW, which are bad for superconductivity, are very different from
the dynamic SDW (see Fig. \ref{fig11}) which is good for superconductivity.
Several possible mechanisms have been proposed to explain these observations
\cite{hoffman,chen0,chenhy,sachdev,podolsky,tesanovic}. It is argued that
the nature of the spin order may not be directly related to the nature of
the charge order, which can be regarded as a result of spin-charge separated
scheme \cite{anderson}. But we consider that a real description of
superconducting electrons should be the charge-\textquotedblleft
spin\textquotedblright -binding picture: charge dimerized and
\textquotedblleft spin\textquotedblright\ coherent. As shown in Fig. \ref%
{fig12}, we have very strong insulating phases which correspond to some
ordering charge \ stripes of simple checkerboard patterns of electron pairs.
This static ordering of stripes may explain why superconductivity disappears
close to these fillings. Moreover the schematic superlattice structure of
Fig. \ref{fig8}, a new charge-\textquotedblleft spin\textquotedblright\
binding picture, suggests that the confining magnetic interactions among the
pairs can also inspire the static SDW and CDW at the same time. As observed
in experiments, the static stripes are in fact in motion \cite{wakimoto}.

\begin{figure}[hp]
\begin{center}
\resizebox{0.85\columnwidth}{!}{
\includegraphics{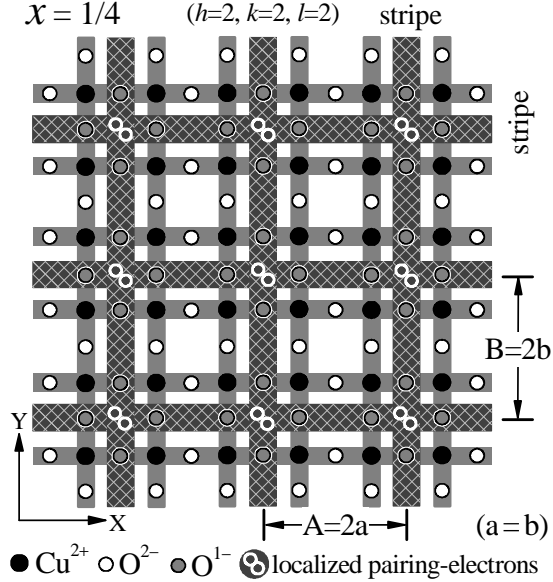}}
\end{center}
\caption{Similar to the underdoped cases of Fig. \protect\ref{fig12}, the
rotationally symmetric charge periodicity of $2a\times 2a$ is inside the CuO$%
_{2}$ plane of the overdoped LSCO at $x=1/4.$}
\label{fig13}
\end{figure}

\subsection{Overdoped ($x=1/4)$}

In contrast to the various and complicated physical phenomena in the
underdoped region, the electronic behavior in the heavy-doped samples
(overdoped) appear to be very simple. It is commonly believed that an
ordering of paired electrons can be formed in the LSCO exhibiting the
anomalous properties in the overdoped region. Spin excitations and transport
properties of LSCO in the overdoped regime have been investigated in detail,
anomalously less-metallic behaviors of the electrical resistivity and the
thermoelectric power have been found in the materials. In particular,
experimental verification of the strong-correlation fluctuations in a
non-superconductive $x=1/4$ sample has been noted \cite{goodenough}. Most
recently, Wakimoto \textit{et al.}\cite{wakimoto1} reported the structural
and neutron-scattering experiment study for over-doped LSCO with $x=1/4.$
They confirmed that the crystal structure of the composition has tetragonal
symmetry (LTT1 phase) with lattice constant of $a=b=3.73$ $\mathring{A}$ at
10 K and the IC peaks appear around the antiferromagnetic wave vector $%
(1/2,1/2$). These interesting observations can be well understood via Eq. (%
\ref{fractions}), which indicates that there exists only one possible LTT1
superlattice phase within this region. As shown in Fig. \ref{fig13}, this
phase has a rotationally symmetric charge periodicity of $2a\times 2a$ at $%
x=1/4$ of $p(2,2,2)$. Accordingly, no superconductivity has been observed at
the $x=1/4$ and even higher doping levels.

It should be noted that our analytical predictions in overdoped region are
also found to be in excellent agreement with the experimental data. This
fact would add considerable support that our theory has the great merit of
explaining high-$T_{c}$ superconductivity.

\begin{figure}[bp]
\begin{center}
\resizebox{0.94\columnwidth}{!}{
\includegraphics{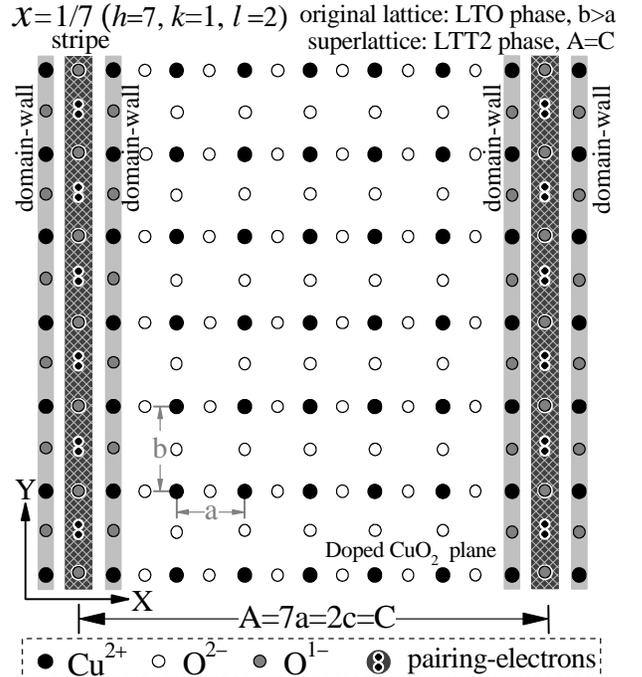}}
\end{center}
\caption{A stable superconducting stripe order in the CuO$_{2}$ plane at $%
x=1/7$, which is near the optimal doping. The charge stripes are segregated
by the domain walls of Cu$^{2+}-$O$^{1-}$ chain. However, unlike Fig.
\protect\ref{fig12} and Fig. \protect\ref{fig13}, here the square
superlattice structure is in the plane perpendicular to the CuO$_{2}$ plane.}
\label{fig14}
\end{figure}

\subsection{Stable superconducting phase ($x=1/7$, 1/14) and optimal doping}

Since the pioneering experiments of Bednorz and M\"{u}ller \cite{bednorz},
the structure and physical properties have been extensively investigated on
the optimally doped LSCO \cite{lee,tsuei2,basov,ino,fischer}. It is commonly
accepted that samples of La$_{2-x}$Sr$_{x}$CuO$_{4}$ have the highest $T_{c}$
at Sr concentration (optimal doping) $x\sim 1.51\ $with the experimental
lattice constants: $a=3.788\mathring{A}$ and $c=13.25\mathring{A}.$ In this
subsection, basing on the Fig. \ref{fig9} and Fig. \ref{fig11}, we will
attempt to provide a general description of the stable superconducting phase
(metallic stripe) in LSCO and give a possible relationship between the
stable doping phase and optimal doping phase.

\begin{figure}[tbp]
\begin{center}
\resizebox{1\columnwidth}{!}{
\includegraphics{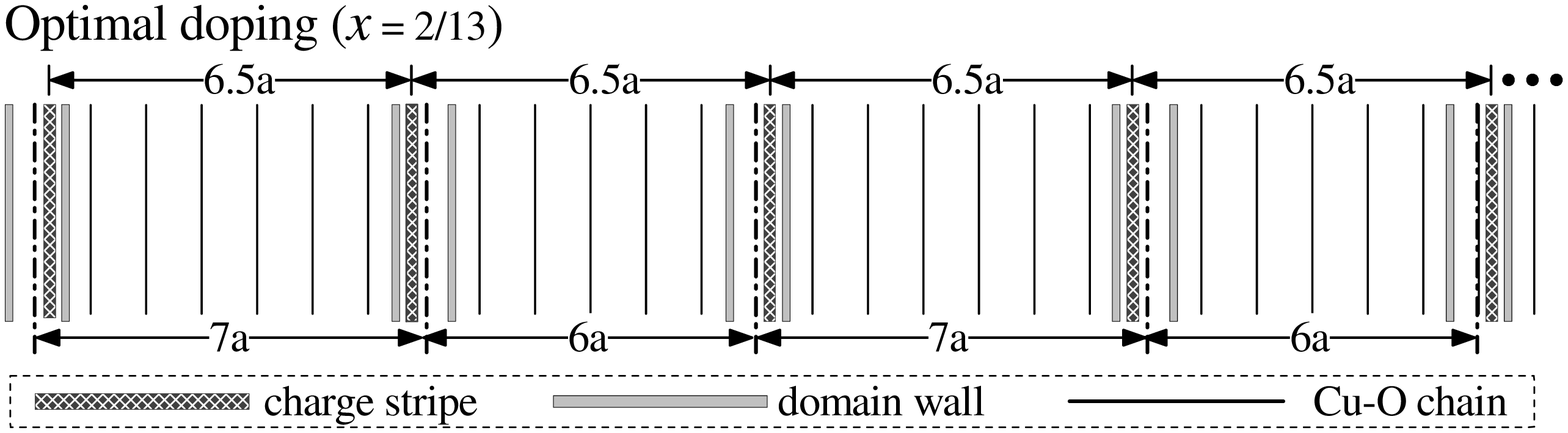}}
\end{center}
\caption{A possible mixed phase in LSCO with the optimal doping $%
x=2/13\approx 1.53$ .}
\label{fig15}
\end{figure}

From these structure parameters, then one has $7a(\sim 26.516\mathring{A}%
)\approx 2c(\sim 26.5\mathring{A})$ which may possibly relate to the
optimally-doped. When $7a=2c$ ($x=1/7\approx 0.143)$, the coexistence of a
stable LTT2 superconductive superlattice\ phase of $p(7,1,2)$ and the LTO
original lattice ($b>a$) of the LSCO is shown in Fig. \ref{fig14}. However,
the analytical value of the optimal doping density ($\sim 0.143)$ is
slightly lower than that obtained through the experimental studies. We argue
that the physically significant critical value for the stable
superconducting phase is not that at which $T_{c}$ is maximum. As shown in
Fig. \ref{fig15}, the maximum high-$T_{c}$ phase may be relevant to a LTT2
related metastable superconducting phase, for example, a uniform mixture of $%
x=1/7$ and $x=1/6$ with an optimal doping at $x\sim 1.53$. In addition,
recall that another possible superconducting stripe (metallic) is shown
schematically in Fig. \ref{fig9}(b), under the restriction of $7a=2c,$ one
can also find another stable superconducting stripe phase at $x=1/14$ of $%
p(7,2,2).$

\section{Phase diagram and superconductive criterion}

\begin{figure}[tbp]
\begin{center}
\resizebox{0.95\columnwidth}{!}{
\includegraphics{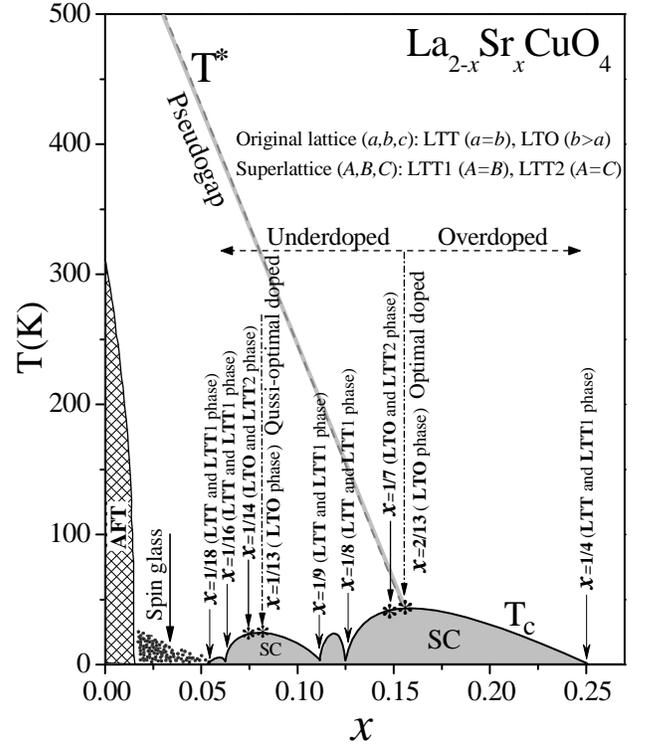}}
\end{center}
\caption{An analytical phase diagram for LSCO. There are five abnormal
phases (at $x=1/4$, 1/8, 1/9, 1/16 and 1/18), where the LTT1 superlattice
phases ($A=B$) can coexist with the LTT original lattices ($a=b$) in the
LSCO.}
\label{fig16}
\end{figure}

For the discussion of our results, we summarize the doping dependence of $%
T_{c}$ for LSCO in a schematic phase diagram in Fig. \ref{fig16}. It is well
known that the antiferromagnetic Mott insulator phase is found near the
origin of La$_{2}$CuO$_{4}$. For doping beyond a few percent, the material
enters the disordered phase (spin glass). At $x=1/18$, the material will
undergo an insulator-to-metal transition, at the same time displaying
superconductivity at low temperature. According to Eq. (\ref{fractions}),
the \textquotedblleft magic effect\textquotedblright\ \cite{moodenbaugh} is
possibly taking place at rational doping levels 1/4, 1/8, 1/9, 1/16 and
1/18, where the LTT1 superlattice phases ($A=B$) can coexist with the LTT
original lattices ($a=b$) in the LSCO. In these specific situations, the
paired electrons are localized, hence the corresponding charge orders appear
to be completely destructive to superconductivity.

We note here that the bosonic theory predicts all magic doping fractions at $%
x=(2m+1)/2^{n},$ where $m$ and $n$ are integers \cite{komiya}, which implies
the possibility of an infinite magic doping fractions in LSCO, while our
theory predicts commensurate effect only at five magic doping fractions 1/4,
1/8, 1/9, 1/16 and 1/18 (see Fig. \ref{fig16}). The reported measurements
find a tendency towards charge ordering at five particular rational doping
fractions of 1/4 \cite{wakimoto1}, 1/8 \cite%
{tranquada,moodenbaugh,crawford,homes,fujita}, 1/9 \cite{zhou1,zhou2}, 1/16
\cite{kim} and 1/18 \cite{wakimoto,keimer,kastner,sugai} and is most
consistent with our theoretical prediction. In view of the intriguing
agreement of the experimental data with our model, it would be desirable to
systematically perform direct measurements of the charge order in the
underdoped LSCO materials, where the nondispersive checkerboard-type
ordering with periodicity $3a\times 3a$ and $6a\times 6a$ can be
experimentally observed at the doping levels $x=1/9$ and 1/18, respectively.

While at $x=1/7$ and $x=1/14$, due to the relation $2c=7a$ in LSCO, the
quasi-one-dimensional metallic charge stripe orders [see Fig. \ref{fig9}(a)
and (b)] can coexist with superconductivity. In particular, our results
imply that the emerging of the superconductivity in high-$T_{c}$ cuprates is
always accompanied by the distortions of the original lattice (a transition
from LTO to LTT phase) in CuO$_{2}$ planes. Thus, the following ratio of the
lattice constants
\begin{equation}
\delta =\frac{b-a}{a},  \label{ratio}
\end{equation}%
where $b\geq a$, can be used to interpret qualitatively the behavior of the
HTSC. When $\delta =0$ (or $b=a$), the superconductivity is totally
suppressed by the stable charge stripes (the square patterns formed inside
the CuO$_{2}$ planes). While $\delta =\delta _{\max }$, the
superconductivity will be enhanced greatly by the anisotropy charge stripes
in the $a-b$ plane and the corresponding value $x$ appears to be the optimal
doping.

\section{Concluding remarks and further experiments}

In conclusion, we have shown a complete replacement of the concept of
quasi-particles (holes) by the real electrons, giving rise to a quite
different electronic pairing and superconductive mechanism. We have found
that two localized electrons can bind together by the photon-mediated
electromagnetic interaction. Our model favors the dominant $d$-wave symmetry
in hole-doped cuprates. In the electron-doped cuprate, the result suggests a
possible mixed $(s+d)$-wave symmetry.

At an appropriate low temperature and doping level, we show that paired
electrons can self-organize into the so-called dimerized Wigner crystal.
Remarkably, two kinds of quasi-one-dimensional metallic magnetic-charge
stripes, where the superconductivity and dynamical spin density wave (SDW)
coexist, have been analytically and uniquely given. Furthermore, the
mechanism has predicted theoretically the \textquotedblleft magic
effect\textquotedblright\ \cite{moodenbaugh} at rational doping levels 1/4,
1/8, 1/9, 1/16 and 1/18 in LSCO. Our results have provided a vivid physical
picture which illustrates clearly how the stripes compete with
superconductivity in high-$T_{c}$ cuprates. Besides, we have presented for
LSCO an analytical phase diagram which is in satisfactory agreement with the
observations from experiments. In fact, the important theoretical \cite%
{anderson0} and experimental \cite{stewart} results imply that the famous
BCS theory \cite{bcs} may be incorrect. Hence, we argue that the any
electronic pairing phenomena should share exactly the same pairing
mechanism. The basic idea of the present paper may also proved to be
important for any other superconductors, for instance, the conventional
superconductors, MgB$_{2}$ \cite{nagamatsu}, and organic superconductors
\cite{williams}.

As is well known since the discovery of the high-$T_{c}$ superconductors, it
is always tempting to construct a universal theory of HTSC which can
naturally explain the complicated problems, such as pairing mechanism,
pairing symmetry, charge stripes, etc. Because of the complexity, the dream
of having such a theory is still a dream. We see that, without Hamiltonian,
without wave function, without quantum field theory, our scenario has
provided a beautiful and consistent picture for describing the myriad
baffling microphenomena which had previously defied explanation. It would be
very significative to recall Anderson's heuristic remark: \textquotedblleft
Many theories about electron pairing in cuprate superconductors may be on
the wrong track\textquotedblright\ \cite{anderson0}. The encouraging
agreement of our results with the experiments implies a possibility that our
theory would finally open a new window in physics. The new ideas presented
in this paper may change the way we view our world.

Finally, are there experiments left to test the conclusions of this paper?
Of course, the answer is yes.

The most direct and convincing test of the theory would be the observation
of the charge orders in the LSCO materials by STM. As shown in Fig. \ref%
{fig16}, our mechanism has definitively determined five abnormal phases (at $%
x=1/4$, 1/8, 1/9, 1/16 and 1/18), where the nondispersive checkerboard-type
orders can coexist with the LTT original lattice ($a=b$) phases of the LSCO.
As we have mentioned in Sec. \ref{magic} that some of the relevant results
have already been reported at doping levels $x=1/4=0.25$ \cite{wakimoto1}, $%
1/8=0.125$ \cite{hanaguri}, and $1/16=0.0625$ \cite{kim}. It is useful to
note that all the three doping parameters are a finite decimal and hence the
corresponding doped samples can be easier to be obtained with high accuracy,
this may explain why the charge periodic superstructures are easy to be
experimentally observed close to these filling. While the other two magic
doping parameters $x=1/9$ $(=0.111\cdots )$ and $x=1/18$ ($=0.0555\cdots )$
are a recurring decimal, as a result, the estimated nondispersive
checkerboard-type ordering with periodicity $3a\times 3a$ and $6a\times 6a$
can only be observed in the LSCO near the doping levels $x=1/9$ and 1/18,
respectively, with much higher accuracy compared to the finite decimal
doping samples.

The neutron scattering experiment can also be used to test the theory. Note
that although the nondispersive $4a\times 4a$ superstructure seems to be
exactly the same in both samples ($x=1/8$ and 1/16) \cite{hanaguri,kim}. We
show, for the first time, that two samples are in fact very different: in
the sample of $x=1/8$ indicated by $p(4,4,1)$ in this paper, where all CuO$%
_{2}$ planes are doped; while at $x=1/16$ of $p(4,4,2)$, only half of the CuO%
$_{2}$ planes (every two planes) are doped. Neutron scattering would be a
way to find the important difference between the \textquotedblleft full
doping\textquotedblright\ phase ($x=1/8$) and the \textquotedblleft half
doping\textquotedblright\ phase ($x=1/16$).

\section*{Acknowledgments}

The author would like to thank {Ron Bourgoin} for many useful discussions
and inspirational remarks.

\end{document}